\documentstyle{article}
\input epsf

\newcommand{\ee}{\mbox{e$^{+}$ e$^{-}$}}
\newcommand{\alphas}{\mbox{$\alpha_s$}}
\newcommand{\alphasM}{\mbox{$\alpha_s(M_{Z})$}}
\newcommand{\sigmanot}{\mbox{$\sigma_{0}$}}
\newcommand{\sinsqchi}{\mbox{$\sin^{2} \chi$}}
\newcommand{\Achi}{\mbox{$ {\cal A}(\chi )$}}
\newcommand{\Bchi}{\mbox{$ {\cal B}(\chi )$}}

\newcommand{\B}{\mbox{$ {\cal B}$}}
\newcommand{\Bcfchi}{\mbox{$ {\cal B}_{C_{F}}\! (\chi )$}}
\newcommand{\Bcachi}{\mbox{$ {\cal B}_{C_{A}}\! (\chi )$}}
\newcommand{\Btrchi}{\mbox{$ {\cal B}_{T_{R}}\! (\chi )$}}
\newcommand{\Bcf}{\mbox{$ {\cal B}_{C_{F}}$}}
\newcommand{\Bca}{\mbox{$ {\cal B}_{C_{A}}$}}
\newcommand{\Btr}{\mbox{$ {\cal B}_{T_{R}}$}}

\newcommand{\Bplus}[1]{\mbox{$ {\cal B}^{+}_{#1}$}}
\newcommand{\Bminus}[1]{\mbox{$ {\cal B}^{-}_{#1}$}}
\newcommand{\Bpm}[1]{\mbox{$ {\cal B}^{\pm}_{#1}$}}

\def\centereps#1#2#3{\vskip#2\relax\centerline{\hbox to#1{\special
  {eps:#3 x=#1, y=#2}\hfil}}}

\def\centerbmp#1#2#3{\vskip#2\relax\centerline{\hbox to#1{\special
  {bmp:#3 x=#1, y=#2}\hfil}}}

\newcommand{\preprint}[1]{\noindent\hfill\hbox{#1}\vskip 10pt}
\newcommand{\writetitle}[3]{
   \vskip 2em
   \begin{center}
        {\LARGE #1 \\}\vskip 1.5em
        {\large \lineskip .5em \begin{tabular}[t]{c} #2 \\
                \end{tabular}\par} \vskip 1em
        {\large #3 } 
   \end{center} 
   \par \vskip 1.5em}

\begin{document}

\title{A Precision Calculation of the Next-to-Leading 
       Order Energy-Energy Correlation Function}

\author{Keith A. Clay and Stephen D. Ellis  \\ 
       {\em {Department of Physics, University of Washington}}}

\date{January 30, 1995}

\preprint{ \vbox{
        \hbox{{\bf UW/PT 94-17}}
        \hbox{{\bf hep-ph/9502223}}  }  }\nopagebreak

\vspace*{8ex}

\writetitle{A Precision Calculation of the Next-to-Leading 
       Order Energy-Energy Correlation Function}
       {Keith A. Clay and Stephen D. Ellis  \\ 
       {\em {Department of Physics, University of Washington}}}
       {January 30, 1995}


\begin{abstract}
The $O(\alphas^2)$ contribution to the Energy-Energy Correlation
function (EEC)~\cite{BBEL,BBELlog,BE,Brown} of 
\mbox{$\ee \rightarrow $ {\em hadrons}} is calculated to high precision and
the results are shown to be larger than previously reported~\cite
{AB,RSE,FK,KN,GS}. The consistency with the leading logarithm approximation
and the accurate cancellation of infrared singularities exhibited by the new
calculation suggest that it is reliable. We offer evidence that the source
of the disagreement with previous results 
lies in the regulation of double singularities.
\end{abstract}

\bigskip

The energy-energy correlation function (EEC)~\cite{BBEL,BBELlog,BE,Brown}
for \ee annihilation into hadrons is widely used as a
measure of the strong coupling constant 
\alphas~\cite{DELPHI,OPAL,SLD} 
and is potentially one of the most precise and detailed experimental
tests of QCD available~\cite{SLD, Bethke}. However, that potential has not
been realized due to disagreement over the predicted value of the
next-to-leading order correction in the strong coupling constant~\cite
{AB,RSE,FK,KN,GS}. We report on a new calculation of the $O(\alphas^2)$ 
term using subtraction for control of infrared singularities. Accuracy
was checked at every stage by symbolic computation, high precision
arithmetic, and human calculation. The detailed cancellation of
singularities in the complicated four-parton states was carefully tested. A
more complete description will be presented elsewhere~\cite{CEnext}.

The EEC was invented to take advantage of the asymptotic freedom of QCD by
viewing the products of \ee annihilation with a weighting
that favored the most energetic hadrons~\cite{BBEL,BE,Brown}. Conservation
of energy requires all energy carried by quarks and gluons to be transferred
to detectable hadrons, hence the EEC is experimentally and theoretically
defined as 
\begin{eqnarray}
\label{eq.def}
\frac{d\Sigma }{d \cos(\chi )} 
        & \equiv &
            \lim_{\stackrel { N \rightarrow \infty }
                {\scriptstyle{\Delta \rightarrow 0}}
                } \;
            \sum_{N \; events}
            \left( \left( \frac{\sigma}{N \, \Delta } \right) 
            \sum_{\stackrel {\scriptstyle{hadrons \; a,b }}
                {\scriptstyle{ \left|
                {\bf \hat{p}}_a \cdot {\bf \hat{p}}_b - \cos(\chi ) \right|
                < \Delta}}}
            \left( \frac{E_a \, E_b}{E_{total}^2}\right) \right)
                 \nonumber \\
        & = &
            \sum_{partons \; i,j} \,
            \int   d^{3}{\bf \vec{p}}_{i} \; d^{3}{\bf \vec{p}}_{j}
            \left( \frac{d\sigma }
                     { d^{3}{\bf \vec{p}}_{i} \, d^{3}{\bf \vec{p}}_{j} } \right)
            \left( \frac{E_i \, E_j}{E_{total}^2} \right)
            \delta \! \left( {\bf \hat{p}}_i \! \cdot \! {\bf \hat{p}}_j 
                               - \cos(\chi )\right) 
\end{eqnarray}
where $\sigma $ is the total cross section for 
\mbox{$\ee \rightarrow $ {\em hadrons}},
$E_n$ and ${\bf \vec{p}}_n$ are the energy and momentum of particle $n$,
and $E_{total}$ is the center of mass energy of the system.  
The EEC is free of collinear
singularities since all parallel momenta are linearly 
summed~\cite{StermanWeinberg}.

After factoring out the trivial dependence on the total cross section and 
\sinsqchi~\cite{KN}, the EEC has the following perturbative
expansion in the region $0<\chi <\pi $, 
\begin{eqnarray}
\label{eq.pert}
 \frac{d\Sigma }{d \cos(\chi )} 
        \equiv &  
                \left(  \frac{ \sigmanot}{ \sinsqchi }
                         \right) 
              \left\{  \left( \frac{ \alphas }{2 \pi } \right)
                 \Achi  \, \left[ 1 + 
                  \left( \frac{ \alphas }{2 \pi } \right)
                        \, \beta_{0} \, 
                        \log \left( \frac{\mu }{E_{total}} \right)
                                \right]  
              +  
                \left( \frac{ \alphas }{2 \pi } \right)^{2} \Bchi +
                O \! \left(  \alphas^3 \right)
                \right\} \, .
\end{eqnarray}
Here \sigmanot\ is the leading order total cross
section, $\mu $ is the renormalization scale, and $\beta _0$ is the leading
coefficient of the $\beta $ function: 
\mbox{$\beta_{0} = \frac{11}{3} C_{A} - \frac{4}{3} T_{R}$}. For QCD in this
notation, \mbox{$ C_{F} = \frac{4}{3}$}, \mbox{$C_{A} = 3$}, and 
\mbox{$T_{R} = \frac{1}{2} N_{F}$}, where $N_F$ is the number of active quark
flavors at energy \mbox{$E_{total}$}. Analytic calculation of 
\mbox{$ {\cal A} $}\ yields~\cite{BBEL} 
\begin{eqnarray}
\label{eq.A}
\Achi & = &
         \; C_{F} \;
         (1 + \omega ) \, (1 + 3 \omega ) 
         \left[ (2 - 6 \omega^{2} ) \log \left( 1 + \omega^{-1} \right)
                + 6 \omega - 3 \right]
\end{eqnarray}
where \mbox{$\omega = \cot^{2} \left( \chi /2 \right)$}. No such analytic
expression is possible for \Bchi. At $O(\alphas^2)$, 
the EEC receives contributions from four-parton final states at tree
level and from three-parton final states with a virtual parton forming one
internal loop. The three-parton final states pose little challenge, but the
integrals corresponding to four-parton states with an external angle fixed
at $\chi $ demand numerical as well as analytic calculation.

\begin{figure}[t]
 \epsffile{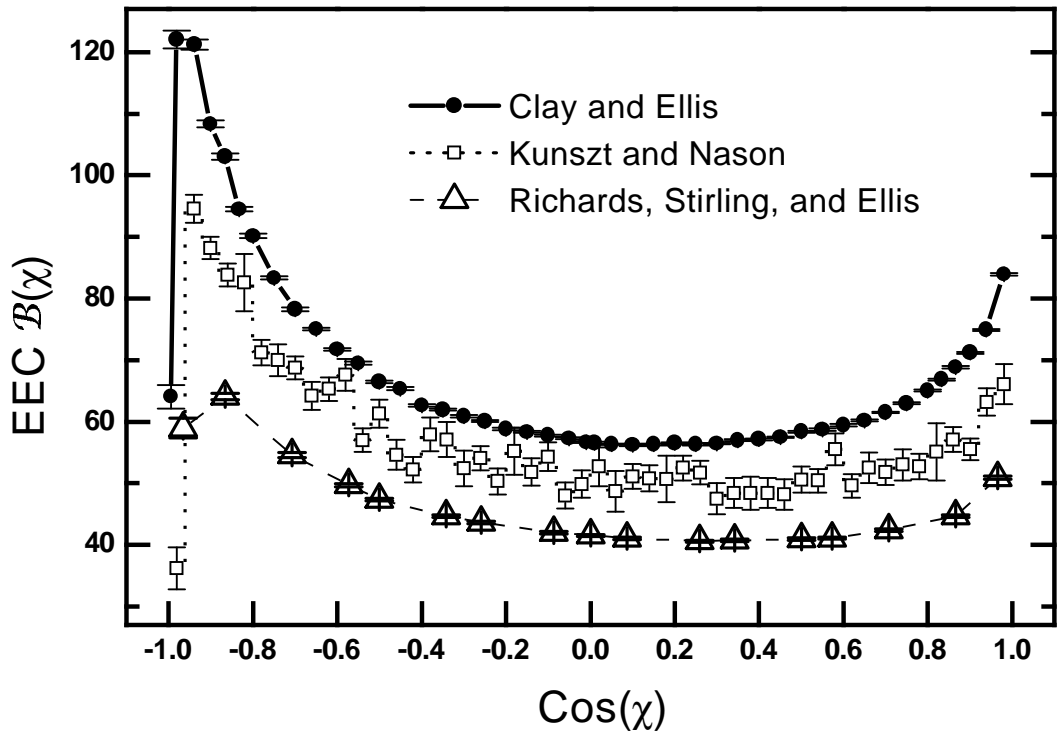}
   \caption{
    \label{fig.b}
    The $O(\alphas^{2})$ contribution to the Energy-Energy Correlation
    function.  For comparison we display our results (solid circles), 
    the results of Kunszt and Nason \protect\cite{KN}  (open squares), 
    and the results of Richards, Stirling, and Ellis \protect\cite{RSE}  
    (open triangles).
    ${\cal B}$ values shown are for five active quark flavors or
    \mbox{$T_{R} = \frac{5}{2}$} (see equation~\protect\ref{eq.B}).}
\end{figure} 

To calculate contributions near soft or collinear poles, the four-parton
expressions were simplified to allow analytic integration in the presence of
an infrared regulator $\epsilon $ (dimension \mbox{$D=4-2\epsilon $}).
Using the subtraction method of infrared regulation,
the simplified expressions were subtracted from exact expressions and
the finite difference was numerically integrated without infrared 
regulation (\mbox{$\epsilon =0$}).  Analytic integrals of the 
three-parton and simplified four-parton expressions 
(at finite $\epsilon$)
were then added and the sum was shown to remain finite in the limit 
\mbox{ $\epsilon \rightarrow 0$}. 
As in all previous calculations of \B, we used the expressions 
derived by Ellis, Ross, and Terrano (ERT)~\cite{ERT} for the exact
three-parton and four-parton final states, but we did not use the ERT 
simplifications or analytic integrals for reasons of maximizing 
numerical convergence.

Our results (Clay and Ellis or CE) are plotted in Figure~\ref{fig.b} along
with the results previously reported by Richards, Stirling, and Ellis 
(RSE)~\cite{RSE} 
and Kunszt and Nason (KN)~\cite{KN}. The mean relative numerical
uncertainty in our calculation is 0.3\%, while for KN it is roughly 4\%,
both arising from the precision of numerical integrations. This
uncertainty is insufficient to explain the roughly 15\% overall difference
between KN and CE. While it is possible for systematic
differences such as these to arise from purely numerical errors, we believe
there is an analytic error at the heart of the disagreement.

\begin{table}[t]
\centering
\begin{tabular}{|c|c|c|c|} \hline \hline \vspace{0.1ex}
Coefficient &   Exact   & Clay and Ellis & Richards, Stirling \\ 
    & Value &    &    and Ellis \\ \hline \hline \vspace{0.1ex}
$\Bplus{3}$ & $ -2 \; C_{F} $ & $ (-2.017 \pm 0.049) \; C_{F}  $ & 
                 $ (-2.46 \pm 0.29) \; C_{F} $ \\ \hline
$\Bplus{2}$ & $ 9 \; C_{F} \; + $ & $ (9.84 \pm 0.90) \; C_{F} \; + $ & 
                 $  (21.0 \pm 9.0) \; C_{F} \; + $ \\ 
    & $ 3.67  \; C_{A} \; + $ & $ (3.63 \pm 0.12) \; C_{A} \; + $ & 
                 $ (2.86 \pm 7.24) \; C_{A} \; + $ \\
    & $ -1.333 \; T_{R}  $ & $ (-1.333 \pm 0.001) \; T_{R}  $ & 
                 $   ( -1.35 \pm 0.05) \; T_{R} $ \\ \hline \vspace{0.1ex}
$\Bplus{1}$ & $ -23.6 \; C_{F} \; + $ & $ (-20.6 \pm 4.79) \; C_{F} \; + $ & 
                 $ (-140 \pm 111) \; C_{F} \; + $ \\
    & $ -1.34 \; C_{A} \; +  $ &  $ (-1.53 \pm 2.11) \; C_{A} \; +  $ &
                 $ (14.0 \pm 71.7) \; C_{A} \; + $ \\
    & $ -0.222 \; T_{R}  $ & $ (-0.220 \pm 0.03) \; T_{R} $ & 
                 $ (-0.066 \pm 0.480) \; T_{R} $ \\ \hline \vspace{0.1ex} 
$\Bplus{0}$ & $ 26.2 \; C_{F} \; + $ & $(23.1 \pm 5.89) \; C_{F} \; +$ & 
                 $ (370 \pm 196) \; C_{F} \; +$ \\
    & $ 16.6 \; C_{A} \; +$ & $(13.43 \pm 9.00) \; C_{A} \; + $ & 
                 $ (-56.8 \pm 228) \; C_{A} \; + $ \\
    & $-3.58 \; T_{R}$ 
             & $(-3.58 \pm 0.17) \; T_{R}$ & 
                 $(-4.16 \pm 1.64) \; T_{R}$ \\ \hline \hline
$\Bminus{1}$ & $ -3.125 \; C_{F} \; + $ & $ (-3.15 \pm 0.04) \; C_{F} \; + $ & 
                 $ (6.51 \pm 0.35)  $ \\
    & $ 3.567 \; C_{A} \; +  $ &  $ (3.57 \pm 0.01) \; C_{A} \; +  $ &
                  ({\em exact} $= 6.533$) $\; +$ \\
    & $ -0.8833 \; T_{R}  $ & $ (-0.8832 \pm 0.0005) \; T_{R} $ & 
                 $ (-0.88 \pm 0.02) \; T_{R} $ \\ \hline \vspace{0.1ex} 
$\Bminus{0}$ &  ?  & $(8.69 \pm 0.40) \; C_{F} \; +$ & 
                 $ 29.9 \pm 2.9 $ \\
    &  ?  & $(15.7 \pm 0.2) \; C_{A} \; + $ & 
                 $ (N_{F} \equiv 4)  $ \\
    & ? & $(-5.46 \pm 0.005) \; T_{R}$ & 
                  \\ \hline \hline
\end{tabular}
\caption{
\label{tab.logs}
The coefficients of the leading log expansion of the EEC at 
large~(\Bplus{j}) and small~(\Bminus{j})
angles. The expansion is as shown in 
equation~\protect\ref{eq.logexp}.  Listed are the exact
leading log coefficients and the coefficients
producing the best fit to Clay and Ellis as well as 
Richards, Stirling, and Ellis \protect\cite{RSElog}.
}
\end{table}

The only known test of the analytic behavior of \B\ is a
comparison with the predictions of the leading logarithm approximation for
large and small angles~\cite{BBELlog}. To determine asymptotic behavior, 
\Bchi\ was calculated over the range 
\mbox{$ \left| \cos(\chi) \right| \leq  (1 - 10^{-6})$}, and the results
were compared to an expansion of the form 
\begin{equation}
\label{eq.logexp}
  \lim _{\eta ^{\pm }\rightarrow 0} \Bchi = C_{F} 
        \sum_{j=0}^{3} \Bpm{j}  \;
        \left[ \ln \!  \left( 1/\eta ^{\pm}\right) \right]^j
\end{equation}
where \mbox{$ \eta^{\pm} = \frac{1}{2} (1 \pm \cos(\chi))$}. The
coefficients \Bpm{j}\ that best fit our calculation
were found using an unconstrained least squares fit and are displayed in
Table~\ref{tab.logs} (we find that 
\mbox{${\cal B}^{-}_{3} = {\cal B}^{-}_{2} = 0$},
as expected). 
For comparison, we also show the coefficients derived
by RSE~\cite{RSElog} who reported some inconsistency with the leading
logarithm approximation.  No inconsistency is evident in our data. The
previously unpublished exact values for \Bplus{0}\ are based on
our conjecture that the form factor for the EEC is the same as that for the
second energy moment of the Drell-Yan cross section~\cite{SWD,SD}. The form
factor is convoluted with a known parton evolution function~\cite{CollSope}
to produce \Bplus{0}.

The discrepancy over the value of
\mbox{$ {\cal B}^{-}_{0}$} is significant.
With $N_{F} \equiv 4$, RSE extracted a value of 
\mbox{$ {\cal B}^{-}_{0}$} equal to
\mbox{$29.9 \pm 2.9$}, while our calculation
predicts a value of \mbox{$47.8 \pm 0.8$}
(see Table~\ref{tab.logs}).
Based on our preliminary analysis of data from KN as well as 
Glover and Sutton (GS)~\cite{GS}, we conclude that neither is 
consistent with the values of \mbox{$ {\cal B}^{-}_{0}$}\ from either 
CE or RSE. It is unfortunate that the coefficient 
that best discriminates between the various calculations is unknown.
An independent calculation of \mbox{$ {\cal B}^{-}_{0}$}\ would be very
useful for resolving the disagreement.

To explore the source of the disagreement, we parameterize 
\B\ as a sum of three functions 
\begin{eqnarray}
\label{eq.B}
\Bchi & = & C_{F} \; \left( \;  C_{F} \; \Bcfchi + 
                                C_{A} \; \Bcachi + 
                                T_{R} \; \Btrchi \; \right)
\end{eqnarray}
and compare our results for each function with those of GS
as well as RSE. While CE and GS~\cite{GSpersonal} 
differ significantly over \Bca\  and 
even more so over \Bcf, they agree with each
other and with RSE~\cite{RSE} on the value of \Btr.
It was also only for 
\Bca\ and \Bcf\ that RSE reported difficulty in the fit to 
leading logarithms~\cite{RSElog}. This strongly suggests that
the source of the disagreement lies outside of the calculation of 
\Btr and is most severely manifest in that of \Bcf.

We believe that the source of disagreement is the regulation of double
(i.e., soft {\em and}\/ collinear) infrared singularities.
Calculation of \mbox{$ {\cal B}_{T_{R}}$}\ involves 
no such regulation since the four-fermion states have no soft 
singularities, while unique to \mbox{$ {\cal B}_{C_{F}}$}\ are
``ladder diagram'' contributions that produce the double singularities 
least controlled by energy weighting.  

To deal with infrared singularities, the exact perturbative integrands 
are simplified in such a way as 
to be analytically integrable in the presence of an infrared
regulator (e.g., $4-2\epsilon$ dimensions) while 
producing integrated expressions that display the same singular 
dependence on the regulator 
(e.g., poles in $\epsilon$) as do integrals of the exact integrands.
The simplified integrands are also used in numerical integrations 
where the regulator is necessarily removed 
(\mbox{$\epsilon \rightarrow 0$}) before integration.
Any such algorithm guarantees that the {\em singular}\/ parts
of the dependence
on the regulator will be correctly calculated.

We have found that simplifications
of integrands involving double poles can produce 
non-singular ($O(\epsilon ^0)$)
errors from inexact treatment of
$O(1/\epsilon )$ shoulders of the $O(1/\epsilon^2)$ double poles
multiplying terms of $O(\epsilon )$.   Since energy weighting can
reposition these shoulders in a complicated way, simplified EEC
integrands may be especially prone to such errors.
These errors cannot be corrected in any 
numerical integrals where 
\mbox{$\epsilon \rightarrow 0$}
prior to integration.  The subtraction method prescribes 
addition and subtraction of the same quantity but 
the added quantities are integrated analytically while 
subtracted quantities must be integrated
numerically to cancel poles in the exact four-parton integrands.
Thus the added and subtracted quantities may differ due to 
necessarily different regulation methods for the numerical 
and analytic integrals.  In such cases, integration of the difference
between simplified and exact integrands is not uniformly convergent
near double poles and the integrals are finite only in the sense
of a numerically computed average.  This average will generally not
be the correct result obtained by analytically setting
\mbox{$\epsilon \rightarrow 0$} after completing integration
rather than before.  

As a test for these errors in our calculation, the cancellation of double
singularities was examined. Since analytic work is difficult for the 
four-parton states, 
we have focused on tests of numerical convergence. The scale
of the independent variable controlling the singularities was magnified by a
factor of $10^4$ in a search for instabilities and neighborhoods of double
poles were divided into separately integrated patches to isolate
divergences. While further study is required, neither test produced signs of
non-uniform convergence or error.

Ultimately theory must be compared with experiment, and fits of our
calculation to data from SLD~\cite{SLD} have been 
performed~\cite{SLDpersonal}. Using the procedure adopted 
in~\cite{SLD}, values for 
\alphasM\ were derived using the EEC as well as the  
asymmetry of the EEC or AEEC:
\begin{displaymath}
   \mbox{$AEEC(\chi) \equiv EEC(\pi - \chi) - EEC(\chi)$}.
\end{displaymath}
Renormalization scales used were in the range  
\begin{displaymath}
     \left. 
        \begin{array}{c}
        {\ 0.0035\;{\textstyle (EEC)}} \\ {\ 0.09\;{\textstyle (AEEC)}} 
        \end{array}
        \right\} \leq \left( \frac{\mu ^2}{E_{total}^2}\right) \leq 4,
\end{displaymath} 
and while fits using KN and CE were found to have similar $\mu $ dependence,
EEC fits using the larger CE values for \B\ yield \alphasM\ values smaller 
by about 0.005~\cite{SLDpersonal,SLDnew}.  Although all 
\B\ calculations yield larger \alphasM\
values from EEC fits than from AEEC fits~\cite{SLD}, it is interesting to
note that the two differ by 0.012 for KN, as opposed to only 0.006 for 
CE~\cite{SLD,SLDpersonal,SLDnew}:
\begin{eqnarray}
  \alphas^{(EEC)}_{(CE)}(M_{Z}) & = &  0.118  
                                \pm  0.013 {\em (scale)} 
                                \pm  0.002 {\em (hadronization)} 
                                \pm  0.003 {\em (experiment)}, 
                \nonumber \\ \nopagebreak
  \alphas^{(AEEC)}_{(CE)}(M_{Z}) & = & 0.112  
                                \pm  0.003 {\em (scale)} 
                                \pm  0.002 {\em (hadronization)} 
                                \pm  0.003 {\em (experiment)} .
\end{eqnarray}
While the improved agreement does not constitute evidence that our
calculation is correct, it is an attractive and suggestive feature of the
results.

We conclude that the disagreement over the next-to-leading order
contribution to the EEC has not been resolved. 
Comparison of our calculation with
all that is known about the EEC shows it to be reasonable and numerically
reliable despite disagreement with previous calculations. A more intensive
investigation of the cancellation of double singularities combined with a
possible extension of our knowledge of the leading logarithm expansion is
needed to resolve the differences.

\bigskip

\begin{description}
\item[Acknowledgments:] \ 

The authors gratefully acknowledge many helpful discussions with
P. Burrows and H. Masuda concerning experimental results from SLD, 
and many useful communications with E.W.N. Glover concerning his 
results.  This research was supported by the U.S. Department of 
Energy, grant number DE-FG06-91ER40614.

\end{description} 

\newpage

\end{document}